\documentstyle[editedbook,epsfig,psfig,epsf]{mq}
\def\etal{{et al.\ }}

\def\cm{{\rm\thinspace cm}}

\def\keV{{\rm\thinspace keV}}

\def\Lsun{\hbox{$\rm\thinspace L_{\odot}$}}

\def\Msun{\hbox{$\rm\thinspace M_{\odot}$}}

\def\ph{{\rm\thinspace ph}}
\def\s{{\rm\thinspace s}}

\def\powerlawfluxat1kev{\hbox{$\ph\cm^{-2}\s^{-1}\keV^{-1}$}}

\def\lapp{\ifmmode\stackrel{<}{_{\sim}}\else$\stackrel{<}{_{\sim}}$\fi}

\def\gapp{\ifmmode\stackrel{>}{_{\sim}}\else$\stackrel{>}{_{\sim}}$\fi}
\def\spose#1{\hbox to 0pt{#1\hss}}

\def\approxlt{\mathrel{\spose{\lower 3pt\hbox{$\sim$}}
        \raise 2.0pt\hbox{$<$}}}
\def\approxgt{\mathrel{\spose{\lower 3pt\hbox{$\sim$}}
        \raise 2.0pt\hbox{$>$}}}

\def\lapp{\ifmmode\stackrel{<}{_{\sim}}\else$\stackrel{<}{_{\sim}}$\fi}
\def\gapp{\ifmmode\stackrel{>}{_{\sim}}\else$\stackrel{>}{_{\sim}}$\fi}

\def\mcg6{MCG$-$6-30-15}
\def\mr2251{MRC~2251-178}
\def\ngc2110{NGC~2110}
\def\13349{IRAS~13349+2438}
\def\iras13349{IRAS~13349}
\def\iras18325{IRAS~18325--5926}
\def\grs1915{GRS~1915+105}
\def\xtej1748{XTE~J1748-288}
\def\chandra{{\it Chandra }}

\def\rxte{{\it RXTE }}

\def\xtegammamcg6{$\Gamma=1.9$}

\def\fe25{Fe~{\sc xxv}}
\def\fe26{Fe~{\sc xxvi}}
\def\Ne9{Ne~{\sc ix }}
\def\ne10{Ne~{\sc x }}
\def\mg11{Mg~{\sc xi }}
\def\si13{Si~{\sc xiii }}

\def\apj{ApJ}

\def\apj{ApJ}

\def\c2{{\sc C~ii}}
\def\c3{{\sc C~iii}]}
\def\c4{{\sc C~iv}}
\def\n5{{\sc N~v}}
\def\o3{[{\sc O~iii}]}

\def\si4{Si~{\sc iv}}
\def\fe25{Fe~{\sc xxv}}
\def\fe26{Fe~{\sc xxvi}}
\def\mg2{Mg~{\sc ii}}
\def\Msun{\ifmmode M_{\odot} \else $M_{\odot}$\fi}
\def\Lsun{\ifmmode L_{\odot} \else $L_{\odot}$\fi}

\def\etal{{\em et al.} }

\def\ergs{erg s$^{-1}$ }

\def\cm2{cm$^2$ }
\def\se1{s$^{-1}$ }



\def\grs1915{GRS~1915+105}
\def\rxte{{\it RXTE }}
\def\chandra{{\it Chandra }}

\begin{opening}
\title{The Chandra HETGS and RXTE view of GRS~1915+105}
\author{J. C Lee$^1$,  C.S. Reynolds$^2$, R. Remillard$^1$, N.S. Schulz$^1$, E.G. Blackman$^3$, A.C. Fabian$^4$}
\institute{$^1$ MIT Center for Space Research, 77 Massachusetts Ave., Cambridge, MA. 02139 U.S.A. \\ $^2$ Dept. of Astronomy, University of Maryland, College Park MD 20742 U.S.A\\ $^3$ Dept. of Physics \& Astronomy, University of Rochester, Rochester, NY 14627  U.S.A.\\ $^4$ Cambridge University -- IoA, Madingley Rd., Cambridge CB3 0HA  U.K.\\ }
\end{opening}
\vspace{-15in}

\runningtitle{The Chandra HETGS and simultaneous RXTE view of \grs1915}
\runningauthor{Lee et al.}

\vspace{-0.5cm}
\begin{document}
\vspace{-0.8cm}
\begin{abstract}
{\small 
The Chandra AO1 HETGS observation of the micro-quasar GRS~1915+105 in
the low hard state reveals (1) neutral K absorption edges from Fe, Si,
Mg, and S in cold gas, and (2) highly
ionized (Fe~{\sc xxv} and Fe~{\sc xxvi}) absorption attributed to a hot disk, disk
wind, or corona. The neutral edges reveal anomalous Si and Fe
abundances which we attribute to surrounding cold material in/near the 
environment of \grs1915.  We also 
point out the exciting possibility for the first astrophysical detection
of XAFS attributed to material in interstellar grains. 
We place constraints on the ionization parameter, temperature,
and hydrogen equivalent number density of the absorber
near the accretion disk based on the detection of the  H- and 
He-like Fe absorption.   Observed spectral changes in the ionized lines 
which track the light curve point to changes in 
both the ionizing flux and density of the absorber, supporting the 
presence of a flow. \\
{\bf Details can be found in Lee et al., 2002, ApJ., 567, 1102 \cite{lee02}. }
}
\end{abstract}

\section{Introduction and Observations}
The {\it Chandra} High Energy Transmission Grating (HETGS) 
and \rxte Proportional Counter Array (PCA) observed \grs1915 
in the low hard state on 2000 April 24 (MJD:
51658.06654, orbital phase zero \cite{ephemeris}) for $\sim$~31.4~ks.
The absolute absorption corrected luminosity $L_{\rm bol} \sim L_X
\approxgt 6.4 \times 10^{38}\ $~\ergs.   The Greenbank Interferometer
radio observations on 2000 April 24.54 indicate a flux of 
$20 \pm 4$ mJy at 2.25 GHz, which is consistent with the presence of a 
steady jet.  

\begin{figure*}[h]
\includegraphics[angle=0,height=1.9in,keepaspectratio=false,width=2.6in]{lee_1.ps}   
\hspace{-0.05in}
\includegraphics[angle=0,height=1.9in,keepaspectratio=false,width=2.6in]{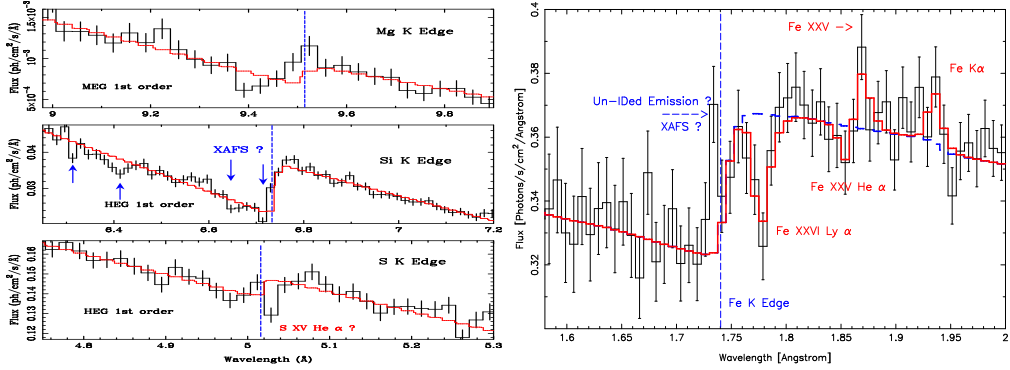}\\
\caption[h]{\footnotesize(LEFT) Photoelectric K-shell edges of S, Si and Mg.   Note the prominence of
the Si edge, and possible XAFS structure.  The Fe~K edge is shown to
the (RIGHT) where ionized (most likely from the accretion disk
atmosphere) H-- and He--like Fe~{\sc xxv} and {\sc xxvi} are also
seen.  Over-plotted is the best fit continuum (dashed) and identified
(solid) lines. The unidentified line may be a shift in the edge due
to XAFS. }
\label{fig-spec}
\end{figure*}

\begin{figure}[b]
\parbox{8truecm}
{\psfig{file=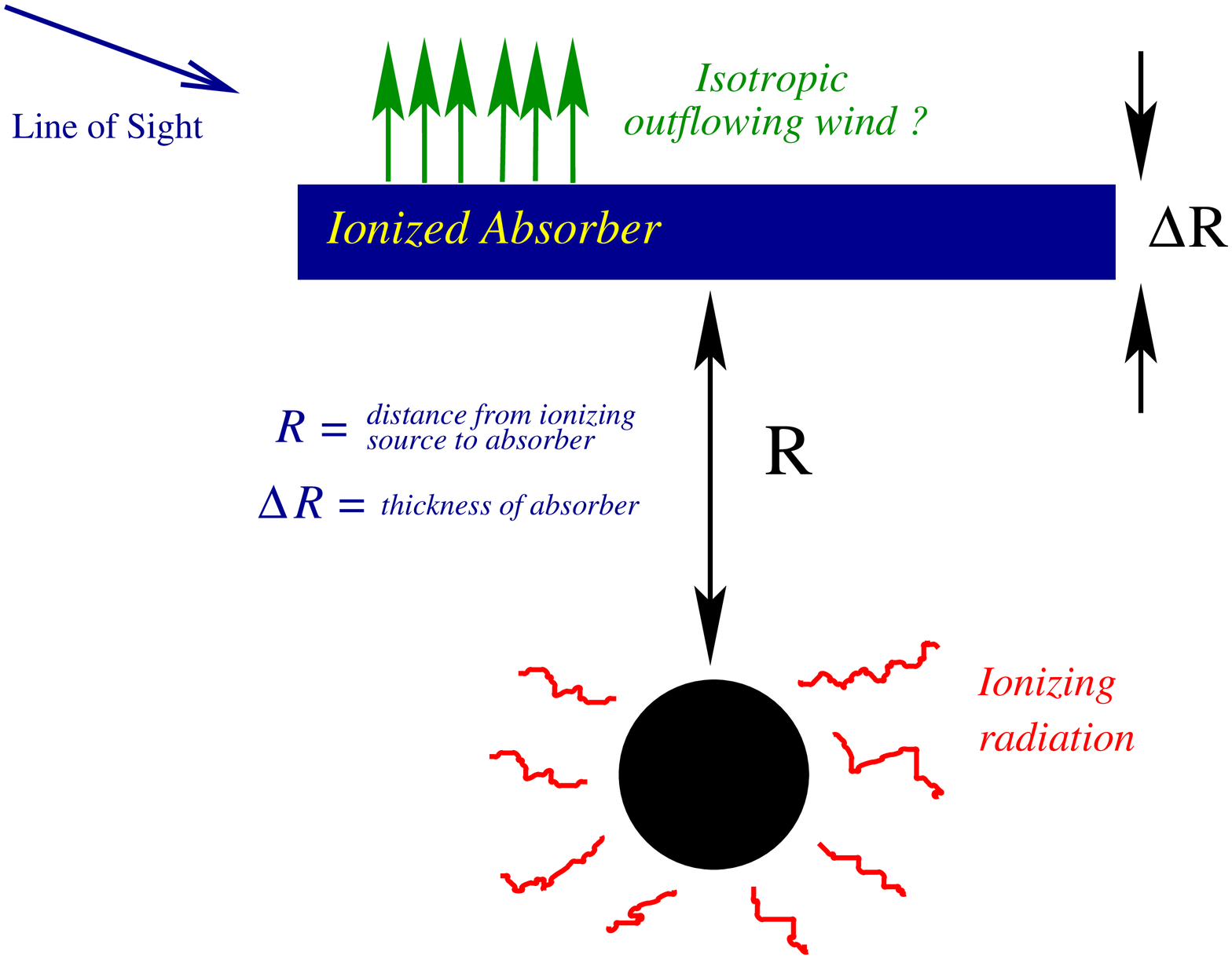,width=8.0cm,height=4.0cm}}
\parbox{5truecm} {{\it Figure 2.}  A possible picture of \grs1915 near the black hole,
as inferred from the \chandra spectra.
For our calculations (\S3), we assume that the volume filling 
factor $\Delta R / R$ must be small
(e.g. $\approx 0.1$)  in order that $\xi$ not change over the
region.  
\label{fig-grs1915}
}
\ \vspace{-0.5truecm}
\  \hspace{0.5truecm}     \
\end{figure}

\vspace{-0.15in}
\section{Surrounding cold material, XAFS and dust}
We detect prominent K-shell absorption edges of Fe, S, Si, and
Mg (Fig.~\ref{fig-spec}). The equivalent Hydrogen column densities 
$N_{\rm H}$ from Mg and S (assuming solar abundances) are consistent 
with the $\sim 3 \times 10^{22} \, \rm cm^{-2}$ 
value expected from neutral line-of-sight absorption.
This is contrasted with the $N_{\rm H} \sim 9 \times 10^{22} \, \rm cm^{-2}$ value
derived from Fe and Si, indicating anomalous abundances, that is likely
associated with the environment of \grs1915.
However, the possibility for unusual line-of-sight abundances cannot 
currently be ruled out. 

We additionally report the exciting possibility for the first astrophysical
detection of XAFS (X-ray Absorption Fine Structure), which is most notable
in the Si edge.   If confirmed, such a positive detection of the details 
of absorption by material in solids (e.g. interstellar grains) will have 
important consequences for (the beginning of) solid state astrophysics,
by which grain properties can be extracted via the solid's inner compound
structure.  

\section{The ionized features from the accretion disk atmosphere}

In addition to the neutral edges, the presence of highly ionized material 
can be surmised from the detection of H-- and He--like Fe~{\sc xxvi}
and Fe~{\sc xxv} resonant absorption and possibly emission.
Assuming photoionization and a $\Gamma = 2$ power law ionizing spectrum, 
we calculate lower limits for the plasma conditions (i.e. ionization parameter $\xi$,
temperature, and equivalent hydrogen column $N_{\rm H}$ in that region) using 
the ratio of the ionic column densities of Fe~{\sc xxvi}~($N_{\rm Fe26} 
\sim 4 \times 10^{17}$) to Fe~{\sc xxv}~($N_{\rm Fe25} \approxlt 1.1 \times 10^{17} \, 
\rm cm^{-2}$)  : 

\[{\bf (I)} \;\;\;\;\; \frac{N_{\rm Fe26}}{N_{\rm Fe25}}  \; \Longrightarrow \;  \left\{ \begin{array}
        {l@{\quad \approxgt \quad \quad}l}
\Huge \rm log \; \xi &  4.15 \, [\rm\, erg\,cm^{-2}\,s^{-1}\,] \;\;\;\;\,  \\
\Huge \rm T & 2.4 \times 10^{\,6} \, \rm K\;\;\;    ({\rm temperature }) \\ 
N_{\rm H} & 2.8 \times 10^{\,22} \, \rm cm^{-2} \;  ({\rm in \;the\; absorber\; region}) \;\;\;\;\;   \\ 
\end{array} \right. \]

Using these results and the relationship between 
the luminosity L and $\xi$,  we can estimate an
upper limit for the distance to the absorber $R$ and a
lower limit to the hydrogen equivalent number density $n$
(see Fig.~2 for illustration).

 \[ {\bf (II)} \;\;\;\;\;{ [\,L \, = \, \frac{\xi}{n\,R^2} \;\; ; \;\; N_{\rm H} = n \, \Delta R\, ]\Longrightarrow \; } \left\{ \begin{array}
        {l@{\quad }lr}
R  \approxlt  & 2 \times 10^{11} \rm\,cm\, \;\;\;\;\;\;\;\;\;\;\;  \rm assumes \\ 
 n \approxgt  & 2 \times 10^{12} \, \rm cm^{-3} \;\;\;\;\;\;\;\;   \rm \frac{\Delta R}{R} \approxlt 0.1 \\ 
\end{array} \right. \] 

\section{Outflowing material and variability}
It is plausible that the ionized absorber (at $R \approxlt 10^{11} \, \rm cm$)
responsible for the H- and He-like Fe lines  is
associated  with material flowing out from the X-ray source
(e.g. wind), which we can compare with the accretion rate.  
We posit a velocity $v \sim 100 \, \rm km\,s^{-1}$ for the calculation of 
the (spherical) mass outflow rate $\dot M_{flow}$ : 
\begin{eqnarray}
\dot M_{flow} =  4\pi r^2 n m_p v \, (\frac{\Omega}{4\pi}) & = & 4\pi m_p v \, (\frac{L_x}{\xi})\,
  (\frac{\Omega}{4\pi}) \sim 9.5 \times 10^{18} \, (\frac{\Omega}{4\pi}) \, \rm gm \, s^{-1}  
\label{eq-Lk}
\end{eqnarray}
where $r$ is the characteristic radius, $\Omega \equiv$~solid  angle subtended 
by the outflow, and the density of the absorbing material $\rho = nm_p$ is 
the product of the proton mass $m_{\rm p}$ and 
the number density of the electrons $n$ in the absorber.
This can be contrasted with
\begin{equation}
\dot M_{accretion} = \frac{L_{bol}}{\eta \, c^2} \sim 7.1 \times 10^{18} \, \rm gm \, s^{-1}
\end{equation}
where the efficiency $\eta \sim 0.1$ and the lower limit on the
bolometric luminosity  $L_{bol} \sim L_x$ is used.  
Such a comparison for $v \sim 100\, \rm km\,s^{-1}$
shows that as the covering fraction (i.e. $\Omega/4\pi$) approaches
unity, the dynamics can be such that the material being accreted is re-released 
in some kind of flow. 

\setcounter{figure}{2}
\begin{figure*}[t]
\includegraphics[angle=0,height=2.0in,keepaspectratio=false,width=2.3in]{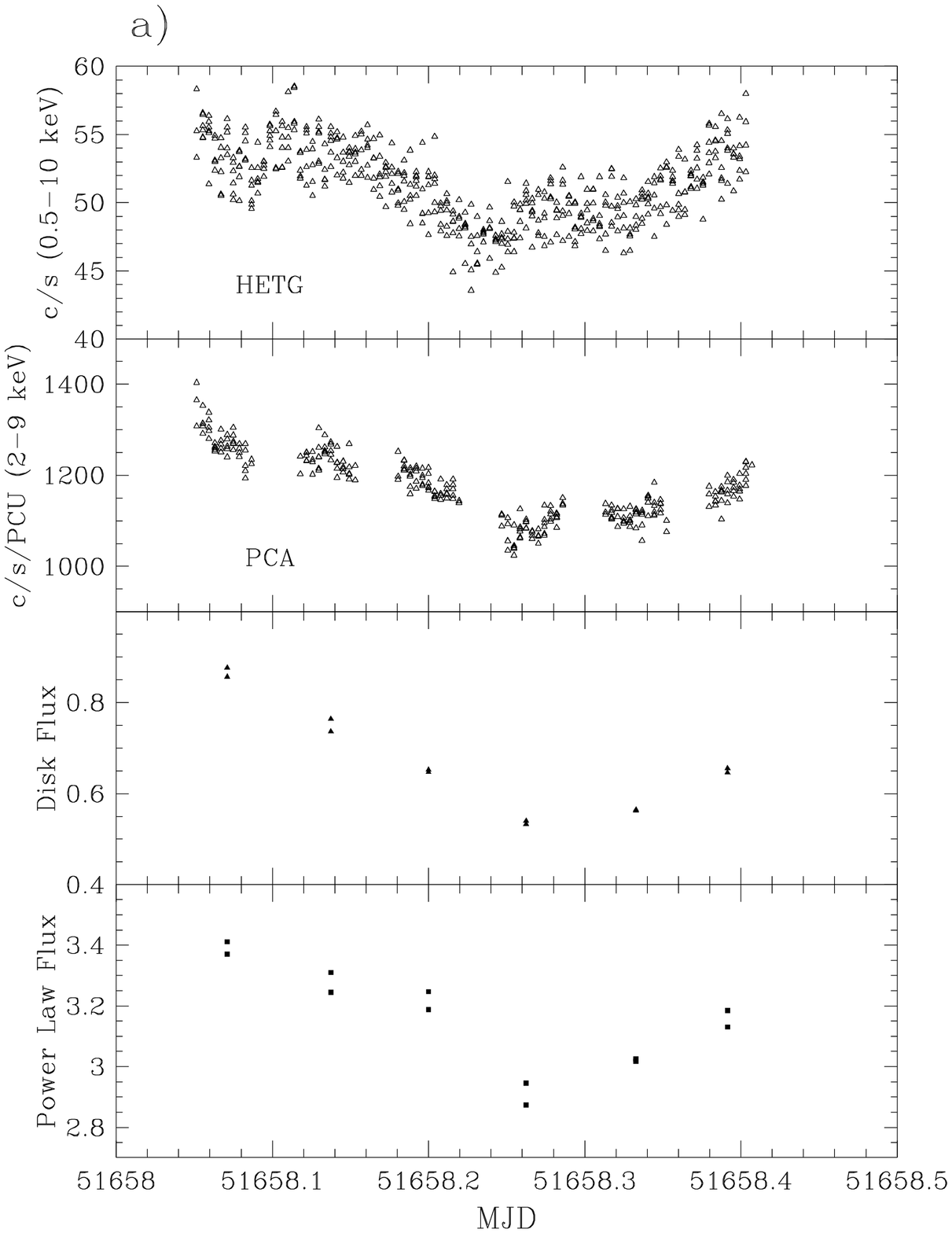}
\hspace{-0.12in}
\includegraphics[angle=0,height=2.0in,keepaspectratio=false,width=3.0in]{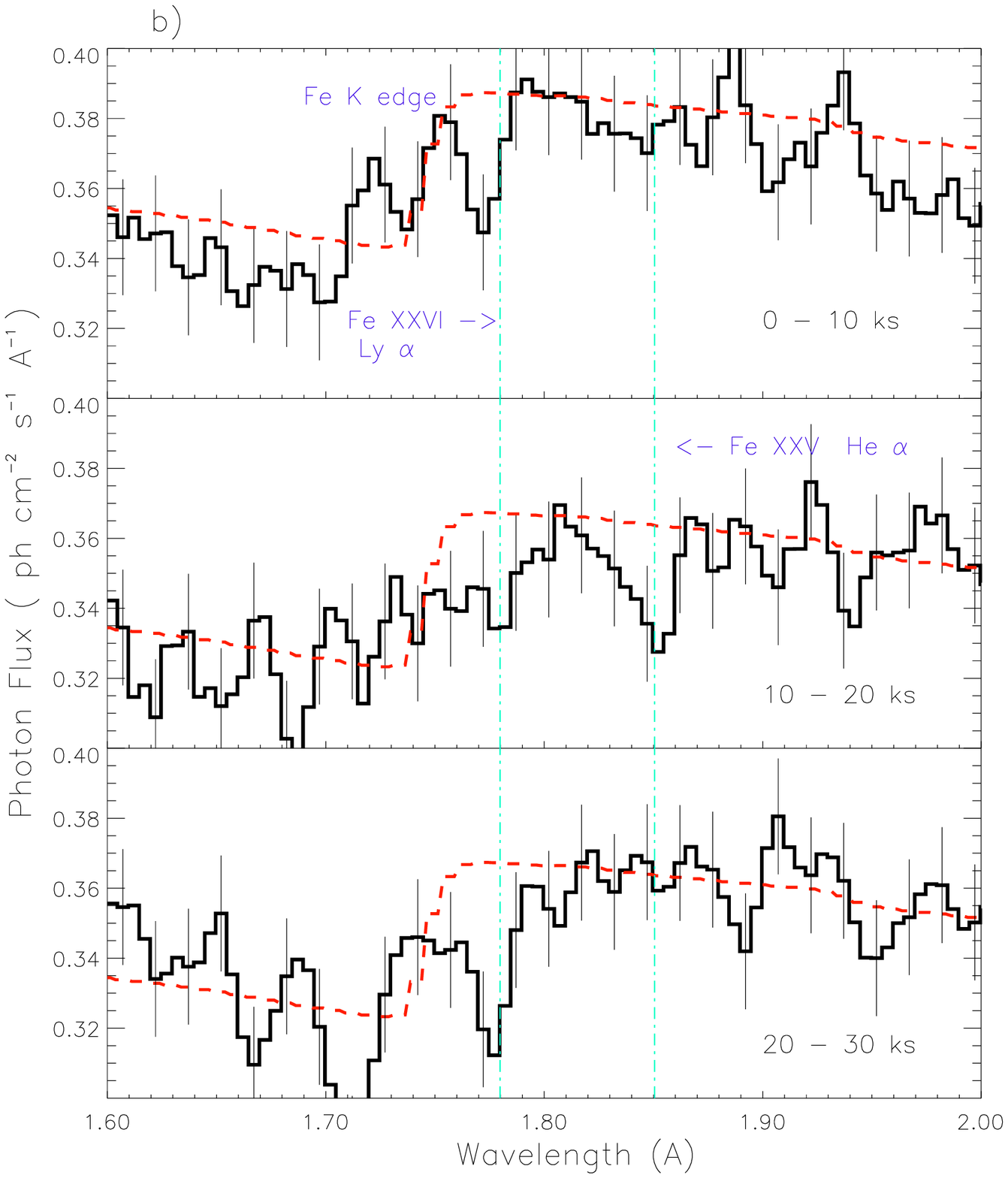}\\
\caption[h]{\footnotesize (a) The \chandra (top panel 1) and \rxte (panel 2) 
light curves during the epoch beginning 2000 April 24 
(MJD: 51658.06654). (Panels 3 \& 4 are in units of
$10^{-8}~\rm erg\, cm^{-2}\, s^{-1}$.)  Changes in the blackbody flux (panel 3) and
power-law flux (panel 4) are coincident with changes  in the light
curve.  (b) The 30~ks \chandra spectrum subdivided into
10~ks intervals reveal the prominence of the Fe~{\sc xxvi} absorption
during the (beginning and end) periods when the flux was highest, in
contrast to the prominence of Fe~{\sc xxv} during the dip.
Error bars are representative of statistically independent bins 
of $\sim 0.0075$~\AA.}
\label{fig-variability}
\end{figure*} 

This possibility for a (slow) outflow is supported by variability results. 
{\bf (1)}~The \rxte data show that the most significant variations occur in the
disk blackbody component which supports an accretion disk origin for the ionized 
lines (Fig.~\ref{fig-variability}a).
{\bf (2)} These variations appear to be correlated with the spectral behavior
seen in the temporal subsets of the \chandra data, 
which show that the ionized lines evolve on $\sim 10$~ks timescales 
(Fig.~\ref{fig-variability}b).  In particular, 
the Fe~{\sc xxv} absorption (Fig.~\ref{fig-variability}b, second panel) 
is more pronounced during the dip in the light curve
(Fig.~\ref{fig-variability}a, panels 1 \& 2), which is also correlated 
with the notable diminution of the disk flux (Fig.~\ref{fig-variability}b,
panel~3).  In contrast, Fe~{\sc xxvi} dominates during the brightest
periods of our observation.  Additionally,  we note that an 
$\sim$~20\% change in the flux at these ionizations  
implies an $\sim$ 30\% change in the ionization fraction of Fe~{\sc xxv}
(assuming that $\xi$ scales with the ionizing flux, while $n$ and $R$ remain constant).
  This is not sufficient to explain the lack of Fe~{\sc xxv} absorption, and
dominance of  Fe~{\sc xxvi} absorption during the periods when the
count rate is high, indicating that the ionizing flux is only part of the solution,  
and changing density the other factor.  For the values of 
$R \sim 10^{11}$~cm and flow velocity $v \sim 10^7 \, \rm cm\,s^{-1}$
discussed earlier, it is plausible that the observed spectral
variability can be attributed to a flow which can change on
timescales of 10~ks.  This is not inconceivable since 10~ks is a very long time, 
corresponding to many dynamical
timescales of the relevant parts of the disk, in the life of a source
like \grs1915.  Furthermore, since this source accretes near  its
Eddington limit, one would expect the mass flow rate and wind density
to be strong functions of the luminosity, such that the actual
structure  of the wind (i.e. density and size) can change as a
function of the luminosity.  

\section{Summary}
The \chandra\, HETGS and simultaneous \rxte\, PCA observations of \grs1915 during the
low hard state reveal : \\

\noindent$\bullet$ {\bf Cold} material with anomalous Fe and Si abundances : 
There is a possibility that these abundance excesses may be related to 
material that is  associated with the immediate environment of \grs1915.
See also similar suggestions from IR studies (\cite{ir1},\cite{ir2}).
Abundance anomalies (of the lighter $\alpha$-process elements such as 
sulfur and oxygen) have also been reported from optical 
spectra of the companion stars of the other microquasars 
GRO~J1655--40 and V4641 Sgr (\cite{m1},\cite{m2},\cite{m3}).
\\

\noindent$\bullet$~{\bf Hot} material from the accretion disk 
atmosphere/wind/corona is detected in the form of highly ionized 
H- and He-like Fe lines. \\
\\
\noindent$\bullet$~{\bf Flow} : The possibility for the presence of 
a slow flow is suggested by variability studies.\\
\\
\noindent$\bullet$~{\bf Dust} :  If confirmed, the detection of XAFS
will have important implications for directly probing the structure and 
chemical composition of interstellar grains. 
\\

\section*{Acknowledgements}
This work was funded by the Chandra grant GO~0-1103X.   JCL also acknowledge
support from the NASA contract NAS~8-01129.

\end{document}